\documentclass[article]{revtex4}
\usepackage{dcolumn,xcolor}

\usepackage{graphicx}% Include figure files
\usepackage{dcolumn}% Align table columns on decimal point
\usepackage{bm}% bold math
\usepackage{amsfonts}
\usepackage{amsmath}
\begin{document}

%\preprint{APS/123-QED}

\title{   Topologically  Stable BPS  and  Non-BPS States in Supersymmetric \(\mathcal{N}=2\) Baby-Skyrme Model }% Force line breaks with \\
%\thanks{A footnote to the article title}%

\author{Emir Syahreza Fadhilla$^{1,2}$}
\email{30221012@mahasiswa.itb.ac.id}
% \altaffiliation[Also at ]{Physics Department, XYZ University.}%Lines break automatically or can be forced with \\
\author{Ardian Nata Atmaja$^2$}%
\email{ardi002@brin.go.id}
\author{Bobby Eka Gunara$^{1,2}$}
\email{bobby@itb.ac.id}
\author{Mir Faizal$^{2,3,4}$}
\email{mirfaizalmir@gmail.com}
\affiliation{%
 $^1$Theoretical High Energy Physics Research Division,\\
 Faculty of Mathematics and Natural Sciences, Institut Teknologi Bandung,\\
 Jl. Ganesha no. 10 Bandung, Indonesia, 40132,
}%
\affiliation{%
 $^2$Research Center for Quantum Physics, National Research and Innovation Agency (BRIN)\\
 Kompleks PUSPIPTEK Serpong, Tangerang 15310, Indonesia,
}
\affiliation{%
$^3$Canadian Quantum Research Center 204-3002 32 Ave Vernon, BC V1T 2L7 Canada.
\\
$^4$Irving K. Barber School of Arts and Sciences, University of British Columbia - Okanagan,\\
Kelowna, BC V1V 1V7, Canada. 
}

%\collaboration{MUSO Collaboration}%\noaffiliation

\date{\today}% It is always \today, today,
             %  but any date may be explicitly specified

\begin{abstract}
  The supersymmetric baby-Skyrme model is an interesting field theoretical model, and its BPS states have been studied using the usual methods. Here,  we propose a novel method to rigorously obtain both topologically stable BPS and non-BPS states in the \(\mathcal{N}=2\) baby Skyrme Model. It is observed that the BPS states found using this novel method coincide with the BPS states found using the usual methods. However, we are also able to obtain the non-BPS states, which break all of the supersymmetry of the theory. Furthermore, there exists a one-parameter family of non-BPS solutions that are connected to the half-BPS solutions, where half of the supersymmetry is restored when the parameter is set to zero. The proposed method is general, and we expect that it might be useful for investigating the topologically stable non-BPS states of other theories. Thus, this method could possibly have wide applications for the study of non-BPS states in supersymmetric theories.  
%\begin{description}
%\item[Usage]
%Secondary publications and information retrieval purposes.
%\item[Structure]
%You may use the \texttt{description} environment to structure your abstract;
%use the optional argument of the \verb+\item+ command to give the category of each item. 
%\end{description}
\end{abstract}

%\keywords{Suggested keywords}%Use showkeys class option if keyword
                              %display desired
\maketitle

%\tableofcontents

\section{\label{Intro}Introduction}
The Skyrme model despite being a simple model, is a physically important model, as it admits topological soliton solutions called Skyrmions \cite{Skyrme:1961vq,Skyrme:1962vh,Brihaye:2005an,Brihaye:2017wqa,Manton:2004tk}. These solutions are connected with  BPS monopoles through a rational map between Riemann spheres \cite{Houghton:1997kg}. 
The proposed model in $(3+1)$ dimensional spacetime is a good low-energy approximation of strongly interacting particles, with the baryon number being related to the topological charge of the Skyrme field \cite{Bolognesi:2006ws, Zahed:1986qz}. 
This model has also been studied holographically using the AdS/CFT correspondence  \cite{Naya:2018mpt, Cartwright:2021tcy}.  
The action of the baby Skyrme model contains all of the possible invariants of strain tensor defined using the map from spacetime to the $S^2$ target space \cite{manton1987,Fadhilla:2021jiz}.
 The Skyrme term is equivalent to the norm-squared topological currents. Since the target space is  $S^2$, the Skyrme field is a $CP(1)$ valued field that can be constructed from the $O(3)$ sigma model multiplets through stereographic projection. 

We can also consider the Skyrme term as a non-standard kinetic term since it is still quadratic in first-derivative with respect to time, but it has higher power factors of first-derivatives with respect to spatial coordinates. The extension of supersymmetry for models that contain non-standard kinetic terms has been actively studied in recent decades (see, for example, \cite{Bazeia:2009db,Adam:2011gc,Adam:2012md,Khoury:2010gb,Khoury:2011da}), and the baby Skyrme model is one of the suitable models to do so because of the Skyrme term contains a  non-standard kinetic term. Thus, it has been possible to generalize the usual baby Skyrme model to a baby  Skyrme model with  $\mathcal{N}=1$  supersymmetry   \cite{Adam:2011hj}. This has been further generalized to a baby Skyrme model with $\mathcal{N}=2$  supersymmetry \cite{Adam:2013awa}. In this study, the BPS states for this $\mathcal{N}=2$   supersymmetric baby Skyrme model have also been obtained using the usual methods based on supersymmetry. It has also been observed using BPS states of a baby Skyrme model that the supersymmetry can be completely restored outside the defect core \cite{Queiruga:2016jqu}. The BPS states of a  $Q$ torus supersymmetric QED have been studied, and it has been demonstrated that they are related to a Skyrmion solution \cite{Bolognesi:2007zz}.  Furthermore, supersymmetric QCD has been demonstrated to have stable Skyrmion solutions because of baryon flat directions \cite{Manohar:1998iy}. Thus, the BPS solutions for the  Skyrme model are physically important. 

We would like to comment that the BPS solutions break some of the supersymmetry of theory \cite{Speziali:2019uzn, Arutyunov:2017dti}. 
Even though it is only possible to have BPS solutions for theories with higher supersymmetry, it has been possible to construct an approximate BPS solution even for a theory with $\mathcal{N} =1$ supersymmetry \cite{Hiller:2002cu}. 
However, it is also possible to have completely non-supersymmetric vacua in string theory \cite{Kaidi:2023tqo, Majumder:1999yy, Honecker:2001dj, Tseytlin:1999ii}.  The conventional BPS methods based on supersymmetry cannot be used to obtain stable soliton solutions for such cases.  Thus, it is important to construct    BPS  solutions using methods that do not rely on the supersymmetry of the theory. The  BPS solutions are obtained as solutions of partial differential equations and depend on the homotopy class of the solution at infinity \cite{Sutcliffe:1997ec}.  
It is the supersymmetry that protects the masses
of states that are destroyed by a linear combination of the supercharges, and so it is due to supersymmetry that these masses become equal to the central charge for BPS states \cite{Gauntlett:2000ch,Gauntlett:1999dt}. However, 
supersymmetry is not essential for constructing stable BPS solutions but is a very convenient technical tool. In fact, BPS solutions for non-supersymmetric theories have been constructed using other methods that do not use supersymmetry \cite{Atmaja:2015umo, Atmaja:2020iti}. Such methods have been used to study the BPS soliton solutions for simple models like the Skyrme model  \cite{Atmaja:2019gce, Fadhilla:2021jiz}. However, such methods can only be applied to very simple theories, and it is not possible to directly apply them to supersymmetric theories, and thus theories that are represented by low-energy effective actions of string theory. 

We would like to point out that one of the motivations to study BPS states in supersymmetric theories comes from string theory, as 
an important and interesting aspect of string theory is the existence of soliton solutions  \cite{Witten:1995im, Duff:1994an, Bagger:2006sk, Konechny:2000dp}. These soliton solutions are localized topologically stable solutions which represent extended geometric objects, and can also interact with other such solutions. 
Due to supersymmetry soliton solutions in string theory are constructed using a special class of states known as BPS states \cite{Witten:1978mh}. The masses of these BPS solutions are equal to their central charge, and their masses do not depend on the couplings. However, apart from 
 from BPS states, string theory also contains stable non-BPS soliton solutions  \cite{Sen:1999md,Sen:1998ii, Kluson:2000iy,Bergman:1998xv}. An example of stable non-BPS states is the $Spin(32)/Z_2$ heterotic string theory \cite{ Gross:1984dd,  Gross:1985rr}. These states in the spinor representation of the gauge group are stable despite being non-BPS states. These states can be obtained using periodic boundary conditions on all the left-moving fermions. They have masses that scale with the string scale, and hence by increasing the string coupling constant, the masses get renormalized. However, states that transform in the spinor representation of $Spin(32)/Z_2$  remain stable. 
These non-BPS solutions can be constructed in certain limits using dualities \cite{Sen:1999md,Sen:1998ii, Kluson:2000iy,Bergman:1998xv}.
The non-BPS black branes in M-theory over Calabi-Yau threefolds have also been investigated using non-BPS black strings in single modulus CICY and THCY models \cite{Marrani:2022jpt}. The Non-BPS fractional branes have been studied in type II string vacua \cite{Sugawara:2018taz}. It has been observed that cylinder partition functions for these fractional branes do not vanish. The Carroll non-BPS D-brane solution have been obtained using the  Carroll limit of a canonical form of unstable D-brane action \cite{Kluson:2017fam}. 
The non-relativistic non-BPS D-brane action has also been constructed, and the properties of the tachyon kink solution on its world-volume have been investigated \cite{Kluson:2006xi}. Thus, non-BPS solutions are physically important in string theory. However,  it has not been possible to use the general formalism such as the supersymmetric BPS formalism, to obtain non-BPS solutions in string theory.  

Furthermore, from a string theory perspective, Skyrme action is important, as it can be used to model D-brane soliton solutions  \cite{Gudnason:2014uha, Prasetyo:2017rij, Gauntlett:2000de, Domokos:2013xqa}.  
It has been proposed that Skyrmions are important in M-theory \cite{Cembranos:2001rp}.
In fact, the Skyrme action has been used for the super-exceptional embedding construction of the heterotic M5-brane \cite{Fiorenza:2020xpx}.  
Among the various Skyrme models that have been constructed, there is a special model in $(2+1)$ dimensional spacetime known as the baby Skyrme model \cite{Piette:1994mh,Piette:1994ug}. The baby Skyrme model has also been used to model a string with $\pi_2$ lump charge \cite{Piette:1994ug, Piette:1994mh}. 
The warped compactification of the two-dimensional
extra space in brane-world models has been investigated using the baby Skyrme action \cite{Delsate:2012hz}. It has been observed that by coupling the fermions with the baby Skyrme, the fermions are localized on the brane \cite{Kodama:2008xm}. The brane-world models constructed using the baby Skyrme action have also been used to model inflation \cite{Brihaye:2010nf, Delsate:2011aa}.
As a baby Skyrme model can be used to model objects in  string theory, and non-BPS states are important in string theory, it is interesting to obtain non-BPS states in a supersymmetric baby Skyrme model. 
So, it is important to analyze not only stable BPS but also non-BPS solutions for  supersymmetric baby Skyrme model. We will obtain such states by generalizing the methods used for calculating the BPS solutions of very simple non-supersymmetric theories \cite{Atmaja:2019gce, Fadhilla:2021jiz} to supersymmetric models. It may be noted that these methods \cite{Atmaja:2019gce, Fadhilla:2021jiz} could not be applied directly to supersymmetric theories, and we had to propose novel methods motivated by them, and it is these novel generalized methods that was applied to the supersymmetric baby Skyrme model.

\section{\label{Method}Novel  Method for both Stable BPS and Non-BPS states} 
In this section, we will propose a novel method that can be used to obtain both BPS and non-BPS states for any supersymmetric theory. It is known that for very simple non-supersymmetric theories, there exists a method called the BPS Lagrangian method, which can be used to obtain  BPS soliton, where the mass does not depend on the coupling constant \cite{Bogomolny:1975de,Atmaja:2015umo}.  The   BPS Lagrangian method,  like the usual Bogomolny formalism \cite{Witten:1995im, Duff:1994an, Bagger:2006sk, Konechny:2000dp},  is based on the quadratic completion of the functionals \cite{Bogomolny:1975de,Atmaja:2015umo}. The difference is that the BPS Lagrangian method focuses on the effective Lagrangian, instead of the static total energy as usually done when we apply the Bogomolny formalism to find the BPS solutions. 
Thus, both methods are equivalent for some cases in the static limit. However, in general, the BPS Lagrangian method possesses a more robust formalism than the Bogomolny formalism since we are dealing with the Lagrangian directly and we can always be sure that the resulting field equations obey the second-order Euler-Lagrange equations.
However, this method has been applied to very simple non-supersymmetric field theories like the ordinary non-supersymmetric  Skyrme model  \cite{Atmaja:2019gce, Fadhilla:2021jiz}.  It has not been generalized to supersymmetric field theories, or applied to string theory backgrounds. Here, we propose that this can be done with some suitable modifications. 

So, in this work, we propose the supersymmetry generalization of the  BPS Lagrangian method \cite{Atmaja:2015umo}. This can be done by first observing  that the  main assumption in the BPS Lagrangian method is that the effective Lagrangian can always be decomposed into quadratic terms plus some boundary terms which consist of the BPS Lagrangian and some remainders, such that
\begin{equation}\label{BPSLagMeth}
    \mathcal{L}_{\text{Eff}}-\mathcal{L}_{BPS}=\sum_{n}K_n\left(\sigma_n\mp G_n\right)^2+\text{remainders},
\end{equation}
where \(K_n(u,\bar{u},X_i(u,\bar{u}),\dots)\) and \(G_n(u,\bar{u},X_i(u,\bar{u}),\dots)\) are functions of the effective fields, \(u,\bar{u},\dots\), and \(\sigma_n(\partial_iu,\partial_i\bar{u},\dots)\) are functions of the derivatives of the effective fields only. The boundary terms, \(\mathcal{L}_{BPS}\) plus the remainder terms, always give trivial Euler-Lagrange equations and are related to the conserved currents of the model (see \cite{Adam:2016ipc} for details). It is worth noting that, in this method, there are freedoms in the choice of the \(\mathcal{L}_{BPS}\) since any modification on the expression of \(\mathcal{L}_{BPS}\) shall only change the resulting expressions of \(K_n\), \(G_n\) and the remainders while keeping \(\mathcal{L}_{\text{Eff}}\) unchanged \cite{Atmaja:2015umo,Atmaja:2018ddi,Atmaja:2019gce}.
If \(\sigma_n\mp G_n=0\) for all \(n\), then these equations are considered as the Bogomolny equations, and the BPS limit $
    \mathcal{L}_{\text{Eff}}=\mathcal{L}_{BPS},
$ 
is achieved if the remainder is equal to zero. As such, the main steps in the BPS Lagrangian method are solving the algebraic equations resulting from the fact that the remainder must be zero at the BPS limit. The earlier versions of \(\sigma_n\) proposed in \cite{Atmaja:2015umo} are polynomials of the derivatives, but it turned out that such functions are not sufficient for some cases, such as the Skyrme models \cite{Atmaja:2019gce}. Thus, \(\sigma_n\) is generalized to any function as long as it only depends on the derivatives of the effective fields. 

A further generalization has also been proposed \cite{Atmaja:2019gce}, where the BPS Lagrangian is not necessarily a sum of boundary terms. This implies that, in general, \(\mathcal{L}_{BPS}\) does not give trivial Euler-Lagrange equations and such equations need to be considered as constraints. This generalization is what sets this novel method apart from original Bogmolonyi's formalism. In the method introduced by Bogomolnyi, the remainder of the quadratic completion, \(\mathcal{L}_{BPS}\), are always boundary terms that are linear to the derivative of the fields, implying that the Euler-Lagrange equations of \(\mathcal{L}_{BPS}\) are always trivial. In contrast with this, the Euler-Lagrange equations of \(\mathcal{L}_{BPS}\) are not trivial and should be considered as constraints for the Bogomolnyi equations acquired from quadratic completion. This generalized approach gives a wider family of solutions and can be used to find not only BPS but also non-BPS solutions. It 
Thus, this method not only reproduces the results obtained using the obtained using Bogmolonyi's formalism but can be used to obtain new results (non-BPS states) that cannot be obtained using the standard Bogmolonyi's formalism. 

It is known that when \(\mathcal{L}_{BPS}\) contains non-boundary terms, the corresponding spatial components of the energy-momentum tensor are non-zero \cite{Atmaja:2018ddi}. In the case of the Skyrme model, these non-zero components imply the instability of BPS solutions \cite{Fadhilla:2020rig,Fadhilla:2021jiz}. 
This improved BPS Lagrangian method has been used to reproduce a wide range of known solutions within the Skyrme model in various spacetime dimensions. Furthermore, the method gives us some new possible solutions, which are either BPS or non-BPS, whose equations are first order in the models where conventional Bogomolny's trick or other known methods cannot be employed.

In this work, we attempt to generalize the BPS Lagrangian method in order to apply it to supersymmetric theories, and thus to string theoretical models. In supersymmetric theory, the bosonic part of the Lagrangian contains terms that explicitly depend on the auxiliary fields, \(F,D,\dots\). This poses a new problem for the BPS Lagrangian method since there are no derivatives of auxiliary fields in the Lagrangian. We have observed that in order to reproduce the on-shell equation from the variational principle on the bosonic sector, Lagrangian with respect to the auxiliary fields, we need to generalize \(\sigma_n\) to be functions of the auxiliary fields as well, i.e. \(\sigma_n\equiv\sigma_n(\partial_i u,\partial_i\bar{u},F,\bar{F},\dots)\). This might seem counter-intuitive at first, but this way we can reproduce the on-shell equation from the Bogomolny equations of the BPS Lagrangian method. The complex norms of the auxiliary fields "shift" the scalar product of the effective fields' derivatives such that they do not show up in the BPS Lagrangian, or at most, push the coefficients of terms that depend on the auxiliary fields to zero. We propose that this is always the case for any supersymmetric models and we are going to demonstrate an example from supersymmetric models with non-standard kinetic terms.

\section{\label{HalfBPSSUSYBabySkyrme} Half BPS States using the Novel Supersymmetric Method  }
We will now demonstrate that the half-BPS states obtained through the proposed supersymmetric generalization of the BPS Lagrangian coincide with the usual BPS method. To show that we will use a supersymmetric baby Skyrme model,  however,  it is important to note that the method is applicable to any supersymmetric theory, and this model is only used as an explicit example. So, we will show that the   BPS states for \(\mathcal{N}=2\) baby Skyrme model obtained using conventional methods \cite{Adam:2013awa}, coincide with the BPS states obtained using the supersymmetric generalization of the BPS Lagrangian method. We start from the \(\mathcal{N}=2\) baby Skyrme model \cite{Adam:2013awa}
\begin{equation}\label{LagBabySkyrme}
    \mathcal{L}=\frac{1}{16}\left[\ln\left(1+\Phi\Phi^\dagger\right)+h(\Phi,\Phi^\dagger)D^\alpha \Phi D_{\alpha}\Phi D^{\dot{\beta}}\bar{\Phi}D_{\dot{\beta}}\bar{\Phi}\right]
\end{equation}
where \(\Phi,\Phi^\dagger\) are the chiral and anti-chiral superfields, and \(D_{\alpha}\) is the \(\mathcal{N}=2\) superderivative (see the Sect. A of Appendix). 
We are interested in static cases, hence, from this point on, we are going to express the equations only in spatial indices, \(i,j,k,\dots\), with \(i,j,k,\dots\in\{1,2\}\). After integration of the Lagrangian \eqref{LagBabySkyrme} with respect to the Grassmann parameters and setting the fermion fields to be trivial, we arrived at the expression for bosonic sector Lagrangian, namely
\begin{eqnarray}\label{LagStatBos}
    \mathcal{L}^{\text{stat}}_{\text{T,bos}}&=&h\left[(\partial_i u\partial^i \Bar{u})^2-\left(i\varepsilon_{jk}\partial_j u\partial_k \Bar{u}\right)^2\right.\nonumber\\
    &&\left.+(2F\Bar{F})\partial_i u\partial^i \Bar{u}+\left(F\Bar{F}\right)^2\right]\nonumber\\
    &&+g\left[\partial_i u\partial^i \Bar{u}+F\Bar{F}\right],
\end{eqnarray}
where \(g\) is the Kaehler metric, \(g=(1+u\bar{u})^{-2}\) for standard baby Skyrme model, and \(h\) is an arbitrary function related to the Skyrme term. The on-shell equation is given by the Euler-Lagrange equation of \eqref{LagStatBos} with respect to \(F\). Let the supersymmetry transformation of the fermion fields expressed by \(\delta \psi=M\epsilon,\) where \(\epsilon\) are Grassmann-valued parameter, then if the eigenvalues of \(M\) are \(\lambda_+,-\lambda_+,\lambda_-,-\lambda_-\), the Bogomolny equation is equivalent to \(\lambda_{\pm}=0\), which implies that the symmetry is preserved. However, since \(F\) and \(u\) must not be trivial then the remaining half of the symmetry must be broken, hence the name, half-BPS states. 

We will now explicitly reproduce the BPS equations obtained through conventional methods \cite{Adam:2013awa}, using the supersymmetric generalization of the  BPS Lagrangian method. If the BPS limit is achievable, then we have \(\mathcal{L}^{\text{stat}}_{\text{T,bos}}=\mathcal{L}_{BPS}\) and the on-shell equation from \(F\) is automatically included in the set of Bogomolny equations. Let \(\gamma\equiv(\partial_i u\partial^i \Bar{u}+F\Bar{F})\) and \(\beta\equiv i\varepsilon_{jk}\partial_j u\partial_k \Bar{u}\) for simplification. According to the supersymmetric BPS Lagrangian method, the minimal BPS Lagrangian consists of linear terms only \cite{Atmaja:2019gce}, given by
\begin{equation}
    \mathcal{L}_{BPS}=-X_1\gamma-Y_1\beta.
\end{equation}
From the BPS limit, we have \(h\gamma^2+\left(X_1+g\right)\gamma-h\beta^2+Y_1\beta=0.\)
Solving this quadratic equation for \((\gamma,\beta)\) gives us the Bogomolny equations and from the remainders, we have \(Y_1^2-(X_1+g)^2=0\).
The remaining constraint can be deduced from the variation of the BPS Lagrangian with respect to the auxiliary field, which gives \(Y_1 = \pm g\).
By substituting this result back to the Bogomolny equations, we managed to reproduce the on-shell BPS equations of the half-BPS states via the BPS Lagrangian method. For the case of Kaehler metric for standard baby Skyrmion, the on-shell BPS equation is given by 
\begin{equation}
    (i\varepsilon_{jk}\partial_j u\partial_k \Bar{u})=\pm\frac{g}{2h}.
\end{equation}
It may be noted that the BPS equations obtained from the conventional method exactly coincide with the BPS equations obtained using this novel method based on the supersymmetric generalization of the BPS method (see Sect. B.1 of Appendix for the conventional method, and Sect. B.2 of Appendix for the novel method).   

In the non-perturbative sector, the total energy for these BPS states can be calculated directly from the Hamiltonian density of static models, namely \(H=- \mathcal{L}^{\text{stat}}_{\text{T,bos}}\). From the Hamiltonian density, we shall proceed by transforming \(u\) into the hedgehog ansatz by assuming that \(\mathcal{L}^{\text{stat}}_{\text{T,bos}}\) is spherically symmetric in \(\mathbb{R}^2\). This can be done by taking \(u=\xi(r)e^{i f(\theta)}\) and both \(g,h\) are functions of \(u\bar{u}\), such that the only possible solution for \(f(\theta)\) is \(f=n\theta\), where \(n\in\mathbb{Z}\) such that \(u\) is periodic in \(\theta\rightarrow \theta+2\pi\), \(n\) is the topological charge of the Skyrmion \cite{Skyrme:1962vh,Manton:2004tk,Atmaja:2019gce}. Substituting the hedgehog ansatz and integrating \(H\) at the BPS limit on the whole space gives us (see Sect. D of Appendix)
\begin{equation}
    E=n\pi.
\end{equation}
The total static energy of the Skyrmion in BPS states is linear in \(n\), as expected, and in general \(n\pi\) is the lower bound of the energy for this supersymmetric baby Skyrme model. Since we know that the topological charge \(n\) is a conserved quantity, it is implied that the energy of the BPS states is topologically protected, hence, such states are stable states. 

%Because the expression of fermion field supersymmetry transformation is not necessary for deriving the BPS equations in this method, it is applicable for more general BPS states where not just one-half of the symmetry is broken. 
\section{\label{NovelBPS}  Non-BPS States using the Novel Supersymmetric Method }
%{\color{red} Physically motivate non-BPS states. Highlight how Usual methods cannot be used to obtain such states.
%Again put most equations in text. }
The non-BPS states represent a wider class of vacua in a supersymmetric theory and it is usually equipped with some free parameters that can be tuned to approximate the BPS states, provided that the non-BPS and the BPS states are connected \cite{Hiller:2002cu}. The usual methods to find the equations for BPS states cannot be used to obtain the first-order equations for non-BPS states. However, the novel methods proposed in this work can be used to obtain not only half-BPS solutions but also non-BPS solutions. To do so, we need to generalize the Lagrangian
\begin{equation}
    \mathcal{L}_{BPS}=-W\beta\gamma-X_1\gamma-X_2\gamma^2-Y_1\beta-Y_2\beta^2-Z,
\end{equation}
\(W,~ X_2,~ Y_2\) and \(Z\) are the additional auxiliary functions. Since the BPS Lagrangian method for this problem only allows, at most, three constraint equations then we choose a particular case that trivially satisfies the constraint from the Euler-Lagrange equation of \(\mathcal{L}_{BPS}\) with respect to \(F\), namely \(X_2=X_1=W=0\) (See Sect. C of Appendix).
Hence, from the BPS limit of this particular case, we have \(
    h\gamma^2+g\gamma+\left(Y_2-h\right)\beta^2+Y_1\beta+Z=0.\)
Solving this quadratic equation gives the Bogomolny equations for \((\gamma,\beta)\) and from the remainders we have \(
0=\left(Y_2-h\right)\left(4Zh-g^2\right)-hY_1^2\).
This leaves us with two functions, \(Y_1\) and \(Y_2\), to be determined by the constraint equation from the variation of \(\mathcal{L}_{BPS}\) with respect to \(u\), hence gives us only one arbitrary functions that parameterize the family of solutions.
If we substitute the Bogomolny equations and the expression for \(Z\) to this constraint, then we have the differential equation for \(Y_1\) that depends on \(Y_2\), namely
\begin{eqnarray}\label{consNovBPS}
        0&=&\left(Y_2-h\right)^3\left(\frac{g^2}{h}\right)'+\left(Y_2'-h'\right)hY_1^2\nonumber\\
        &&+\frac{(Y_2-h)(Y_2-2h)}{2}\left(Y_1^2\right)',
\end{eqnarray}
with \((.)'\) denotes the derivatives with respect to \(u\bar{u}\).
By relating this constraint to the supersymmetry eigenvalues, we conclude that half of the symmetry is preserved if and only if \(Y_2=0\) or \(Y_2=h/g\). Since for general \(Y_2\) all of the supersymmetries are broken, then the solutions with \(Y_2\neq 0\) correspond to the non-BPS states. The arbitrary \(Y_2\) should be understood as the quantity which determines the deformation of the vacua from BPS to the non-BPS ones.

In contrast with the BPS states of the supersymmetry extended model discussed in section \ref{HalfBPSSUSYBabySkyrme} where the term in the RHS entering the BPS equations is strictly constrained by the Kaehler metric, the solutions of this completely broken state do not depend on the Kaehler metric due to the freedom on the choice of \(Y_2\) mentioned earlier. Instead, we should choose a specific "potential" term for the BPS equations and a specific Kaehler manifold as well in order to find a unique \(Y_2\), hence we can state that \(Y_2\) depends on the choice of \(g\) for specific BPS states.

%When \(Y_2\neq 0\) then we shall expect more broken symmetry. Let us consider the quarter(\(1/4\))-BPS states where three out of four fermion transformation is broken. \dots

Let us consider a special case where \(Y_2\) is linear to \(h\), namely \(Y_2=C_1 h\) with \(C_1\) is a real constant, and \(h\) satisfies \(h=g^2\). Such states should, in general, break all supersymmetries. At this point, \(C_1\) is still arbitrary but its value should be constrained by the behaviour of the solution for field \(u\) and the total energy, such that the corresponding vacua are well-behaved and obey the BPS bound. From equation \eqref{consNovBPS} we deduce that
for this special case, the on-shell Bogomolny equation is given by 
\begin{equation}
    i\varepsilon_{jk}\partial_j u\partial_k \Bar{u}=\pm\frac{1}{\sqrt{2}}\left[\frac{(2-C_1)}{(1+u\bar{u})^4}\right]^{-\frac{(C_1-1)}{(C_1-2)}}.
\end{equation}
The Bogomolny equation shown above is connected to the BPS equation by taking the limit \(C_1=0\).

There are some interesting properties of the non-BPS states. In general, solutions of \(u\) in closed form are not possible, hence we assume the hedgehog ansatz that we have used for the BPS states. From here, calculating the total energy is straightforward (see Sect. D of Appendix), 
\begin{eqnarray}\label{energy}
    E&=&n\pi\sqrt{2}(2-C_1)^{-\frac{C_1-1}{C_1-2}}\frac{\Gamma\left(\frac{2+C_1}{2-C_1}\right)}{\Gamma\left(\frac{4}{2-C_1}\right)}.
\end{eqnarray}
We can see that the energy is linear in \(n\), implying that the BPS property \(E(n_1)+E(n_2)=E(n_1+n_2)\) is still preserved for these solutions although all the supersymmetries are broken. As such, these states are stable because they are topologically protected states unless \(n=0\). Furthermore, from the expression of total energy \eqref{energy}, we know that \(C_1\in(-2,0]\) because, outside this interval, the energy should be positive definite and higher than the BPS bound, \(n\pi\).

\begin{figure}[h]
    \centering
    \includegraphics[width=0.65\textwidth]{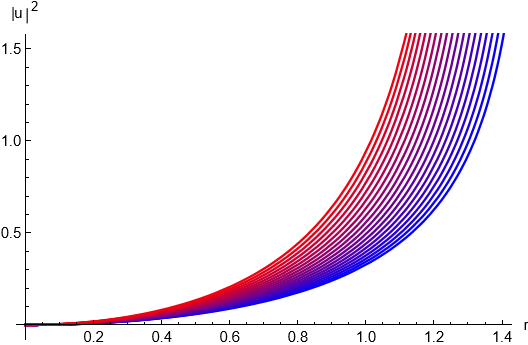}
    \caption{Plot of \(u\Bar{u}\) against \(r\) with \(n=1\) for varying \(-1.9\leq C_1\leq 0\) with increments of \(0.1\) form the positive branch. The red curves are solutions with smaller \(C_1\) increasing to the bluer ones with \(C_1\) getting close to zero.} 
    \label{fig:uUPlot}
\end{figure}

There are two possible sets of solutions which we will call the negative branch and positive branch. \(|u|^2\) of the negative branch solutions are monotonically decreasing in \(r\) and are always negative, hence we should omit the negative branch. For the positive branch, there is a cut-off radius
\begin{equation}
    r_{\text{cut-off}}=\frac{2^{3/4} (2-C_1)^{\frac{1}{2}
   \left(2-\frac{1}{2-C_1}\right)}
   \sqrt{n}}{\sqrt{2-3 C_1}},
\end{equation}
implying that they are possibly compactons. The influence of the parameter \(C_1\) on the compacton size is demonstrated in figure \ref{fig:uUPlot}.

The topological charge density, given by \(q=(n/\pi)\xi\xi'/r(1+\xi^2)^2,\) can be used to ensure whether the solutions with cut-off radius are indeed compactons solutions or just non-physical solutions. We found that for  \(-2\leq C_1\leq0\) the topological charge density vanishes at the cut-off radius since \(\xi^2=|u|^2\rightarrow\infty\) at \(r_{\text{cut-off}}\). This implies that \(q\rightarrow\) at finite \(r\) which implies that the solutions are compactons with compact support \(r\in[0,r_{\text{cut-off}}]\).

%In figure \ref{fig:compacton}, we show the distribution of topological charge density for both the BPS and non-BPS vacua.
%\begin{figure}
%    \centering
%    \includegraphics[width=0.45\textwidth]{TopologicalCharge.png}
%    \includegraphics[width=0.45\textwidth]{TopologicalCharge1.png}
%    \caption{Topological charge density for \(C_1=0.6\) (left) and for \(C_1=1\) (right)}
%        \label{fig:compacton}
%\end{figure}

\section{Conclusions and Outlook}
We have proposed a supersymmetric generalization of the  BPS Lagrangian method that can be used to find   BPS and non-BPS solutions in supersymmetric theories. The important observation is that \(\sigma_n\)'s in equation \eqref{BPSLagMeth} need to be functions of, not just derivatives of the effective fields, but also the auxiliary fields as well. We believe that this is due to the fact that auxiliary fields do not enter the BPS Lagrangian in order to have on-shell BPS equations. The method is also proven to be applicable for finding BPS states of models where usual techniques, such as Bogomolny's formalism or supersymmetry transformation eigenvalues, cannot be employed.  

For the baby Skyrme model \cite{Adam:2013awa}, we managed to find novel non-BPS states with Bogomolny equations whose family of solutions is parametrized by an arbitrary function, \(Y_2\). These states have non-zero eigenvalues, hence all the supersymmetry is broken but the auxiliary field, \(F\), still satisfies the on-shell equation and the half-BPS states are recovered if and only if \(Y_2=0\). All of the on-shell BPS equations found resemble the BPS equation of the original BPS baby Skyrme model where the "potential" term enters the equation as a function of not just Kaehler metric as expected, but also as some arbitrary functions. Thus, we brought back the freedom to choose the "potential" for the BPS states for any Kaehler manifolds by breaking all the supersymmetry as a consequence.

We would also like to point out that even though we have explicitly applied them to a supersymmetric baby Skyrme model, the novel methods proposed here are general and could potentially be useful to study topologically stable BPS and non-BPS solitons in other supersymmetric theories. Hence, it could be possible to study the implication of this method to other supersymmetric models in string theory. In fact, one of the important aspects of this work is that we were able to obtain stable BPS   solutions without using the standard formalism based on supersymmetry.  As this method is general, it could have potential applications to other backgrounds in string theory, where all the supersymmetry is broken and standard BPS methods based on supersymmetry cannot be used. Hence, this work could be used to obtain general non-BPS soliton solutions, and it could also be used to study non-BPS solutions in string theory. Hence, it would be interesting to apply this method to study topologically stable non-BPS solutions in other field theoretical models motivated by string theory.

\begin{acknowledgments}
The work in this paper is supported by GTA Research Group ITB and ITB Research Grant. M.F. acknowledges the hospitality and support from BRIN, where this work was carried out. 
B. E. G. acknowledges the financial support through BRIN Visiting Researcher Programme. E. S. F. also would like to acknowledge the support
from BRIN through the Research Assistant Programme 2023.
\end{acknowledgments}
\bibliography{apssamp}% Produces the bibliography via BibTeX.
\appendix
\section{Supersymmetric \(\mathcal{N}=2\) Baby Skyrme Model}
In this section, we first describe the supersymmetry extension of the Baby Skyrme model which was first proposed in [58, 59]. 
Let \(\theta^\alpha\) and \(\bar{\theta}^{\dot{\alpha}}\), \(\alpha\in{1,2}\), be Grassmann variables and the corresponding supersymmetry generators \((Q,\bar{Q})\) satisfying the algebra
\begin{equation}
    \{Q_\alpha,\bar{Q}_{\dot{\alpha}}\}=-2i\sigma^\mu_{\alpha\dot{\alpha}}\partial_\mu,~~~\{Q_\alpha,Q_{\beta}\}=\{\bar{Q}_{\dot{\alpha}},\bar{Q}_{\dot{\beta}}\}=0.
    \end{equation}
can be expressed as
\begin{equation}
    Q_\alpha=\frac{\partial}{\partial \theta^\alpha}-i\sigma^\mu_{\alpha\dot{\alpha}}\bar{\theta}^{\dot{\alpha}}\partial_\mu,~~~
    \bar{Q}_{\dot{\alpha}}=\frac{\partial}{\partial \bar{\theta}^{\dot{\alpha}}}-i\theta_\alpha\sigma^\mu_{\alpha\dot{\alpha}}\partial_\mu.
\end{equation}
The superderivatives \((D_\alpha,\bar{D}_{\dot{\alpha}})\), where
\begin{equation}
    D_{\alpha}=\frac{\partial}{\partial\theta^\alpha}+i\sigma^\mu_{\alpha\dot{\alpha}}\bar{\theta}^{\dot{\alpha}}\partial_\mu,~~
    \bar{D}_{\dot{\alpha}}=-\frac{\partial}{\partial\bar{\theta}^{\dot{\alpha}}}-i\theta^\alpha\sigma^\mu_{\alpha\dot{\alpha}}\partial_\mu,
\end{equation}
satisfy similar anti-commutation relations as \((Q,\bar{Q})\),
\begin{equation}
    \{D_\alpha,\bar{D}_{\dot{\alpha}}\}=-2i\sigma^\mu_{\alpha\dot{\alpha}}\partial_\mu,~~~\{D_\alpha,D_{\beta}\}=\{\bar{D}_{\dot{\alpha}},\bar{D}_{\dot{\beta}}\}=0.
\end{equation}
and both set anti-commute,
\begin{equation}
    \{D_\alpha,Q_\beta\}=\{\bar{D}_{\dot{\alpha}},\bar{Q}_{\dot{\beta}}\}=\{\bar{D}_{\dot{\alpha}},Q_\beta\}=\{D_\alpha,\bar{Q}_{\dot{\beta}}\}=0.
\end{equation}
We can construct the $\mathcal{N} =2$ chiral superfield and its corresponding anti-chiral superfield from the conditions
\begin{equation}
    \bar{D}_{\dot{\alpha}} \Phi=0,~~~D_{\alpha}\Phi^\dagger=0.\label{ChiralCond}
\end{equation}
According to [59] the expression of superfield and its anti-chiral counterpart which satisfy the chiral conditions are given by
\begin{eqnarray}
    \Phi&=&u+i\theta\sigma^\mu\bar{\theta}\partial_\mu u+\frac{1}{4}\theta\theta\bar{\theta}\bar{\theta}\partial^\mu\partial_\mu u+\sqrt{2}\theta\psi-\frac{i}{\sqrt{2}}\theta\theta\partial_\mu\psi\sigma^\mu\bar{\theta}+\theta\theta F\\
    \Phi^\dagger&=&\bar{u}-i\theta\sigma^\mu\bar{\theta}\partial_\mu \bar{u}+\frac{1}{4}\theta\theta\bar{\theta}\bar{\theta}\partial^\mu\partial_\mu \bar{u}+\sqrt{2}\bar{\theta}\bar{\psi}+\frac{i}{\sqrt{2}}\theta\sigma^\mu\partial_\mu\bar{\psi}\bar{\theta}\bar{\theta}+\bar{\theta}\bar{\theta}\bar{F},
\end{eqnarray}
where \(\psi\) is the fermion field, \(F\) is the complex-valued auxiliary field, and \(u\) is the complex function that parametrizes the target space of Skyrme field through \(CP(1)\) formulation. As such,
the Lagrangian of $\mathcal{N} =2$ SUSY extended baby Skyrme model consist of two terms, namely the quadratic term, \(\mathcal{L}_2\), and the quartic term, \(\mathcal{L}_4\), such that it can be expressed as
\begin{equation}\label{SUSYLag}
    \mathcal{L}=\mathcal{L}_2+\mathcal{L}_4.
\end{equation}
The quadratic and quartic terms themselves are constructed such that their bosonic sector resembles the quadratic and quartic terms of the non-SUSY baby Skyrme model, namely
\begin{eqnarray}
    \mathcal{L}_2=\frac{1}{16}\ln\left(1+\Phi\Phi^\dagger\right),~~~\mathcal{L}_4=\frac{h(\Phi,\Phi^\dagger)}{16}D^\alpha \Phi D_{\alpha}\Phi D^{\dot{\beta}}\bar{\Phi}D_{\dot{\beta}}\bar{\Phi},
\end{eqnarray}
with \(h(\Phi,\Phi^\dagger)\) is an arbitrary function of the chiral superfields that generates the factor of the quartic term of \(CP(1)\) baby Skyrmion, for example, \(1/(1+u\bar{u})^4\) for the standard baby Skyrme model. The model given above has a Kaehler metric \(g=1/(1+u\bar{u})^2\). By integrating the Lagrangian above with respect to all the Grassmann-valued parameters, \(\theta\) and \(\bar{\theta}\), we can find the expression of the bosonic sector Lagrangian of expression \eqref{SUSYLag}, which is given by
\begin{eqnarray}\label{LTBos}
    \mathcal{L}_{\text{T,bos}}&=&g\left[\partial_\mu u\partial^\mu \Bar{u}+F\Bar{F}\right]+h\partial_\mu u\partial^\mu u\partial_\nu \Bar{u}\partial^\nu \Bar{u}+2hF\Bar{F}\partial_\mu u\partial^\mu \Bar{u}+h\left(F\Bar{F}\right)^2,
\end{eqnarray}
where \(g\equiv g(u,\Bar{u})\), \(h\equiv h(u,\Bar{u})\), \(u\) is the Skyrme field, and \(F\) is the auxiliary field. By assuming that all of the fermion fields are trivial, the non-trivial on-shell states can be obtained from Euler-Lagrange equations of \eqref{LTBos} with respect to both \(u\) and the auxiliary field \(F\).

\section{Bogomolny Equations of the Half-BPS states}
\subsection{The Standard Method from Eigenvalues of SUSY Transformations}
The standard way to find the half-BPS solutions is by considering the SUSY transformation and the transformation is trivial if it is a SUSY-preserving state. We need to impose the condition that \(u\) and \(F\) must not be trivial in order to have a solution that leads to a possibility that the transformation of the fermion does not entirely preserve the SUSY. For our static case, the fermion transformation is given by
\begin{eqnarray}
    \delta\psi_1&=&\partial_1u\Bar{\epsilon}^{\dot{1}}-\partial_2u\Bar{\epsilon}^{\dot{2}}+F\epsilon_1,\\
    \delta\psi_2&=&-\partial_1u\Bar{\epsilon}^{\dot{2}}-\partial_2u\Bar{\epsilon}^{\dot{1}}+F\epsilon_2,\\
    \delta\psi^{\dot{1}}&=&\partial_1\Bar{u}\epsilon_1-\partial_2\bar{u}\epsilon_2+\Bar{F}\Bar{\epsilon}^{\dot{1}},\\
    \delta\psi^{\dot{2}}&=&-\partial_1\Bar{u}\epsilon_2-\partial_2\bar{u}\epsilon_1+\Bar{F}\Bar{\epsilon}^{\dot{2}}.
\end{eqnarray}
If we recast the set of transformations above into
\begin{equation}
    \delta \psi=M\epsilon,
\end{equation}
with \(\delta \psi=M\epsilon,\) with \(\delta\psi=(\delta\psi_1,\delta\psi_2,\delta\bar{\psi}^{\dot{1}},\delta\bar{\psi}^{\dot{2}})\), and \(\epsilon=(\epsilon_1,\epsilon_2,\bar{\epsilon}^{\dot{1}},\bar{\epsilon}^{\dot{2}})\) where \((\epsilon_i,\Bar{\epsilon}^{\dot{i}})\) are Grassman-valued transformation parameters, then the eigenvalues of the transformation matrix, \(M\), are \(\lambda_+,-\lambda_+,\lambda_-,-\lambda_-\) and these eigenvalues of supersymmetry transformation on fermions satisfy
\begin{equation}
    \lambda_{\pm}^2=-\left(\partial_i u\partial^i \Bar{u}+F\Bar{F}\right)\pm i\varepsilon_{jk}\partial_j u\partial_k \Bar{u}.
\end{equation}
One should get the Bogomolnyi equations when one of the symmetries is preserved by taking \(\lambda_{\pm}^2=0\), which gives
\begin{equation}\label{Equ}
    \partial_i u\partial^i \Bar{u}+F\Bar{F}=\pm i\varepsilon_{jk}\partial_j u\partial_k \Bar{u},
\end{equation}
and, together with the on-shell equation, 
\begin{equation}\label{EqF}
    F\bar{F}=-\partial_i u\partial^i \Bar{u}-\frac{g}{2h},
\end{equation}
these are the BPS equations for one-half BPS states [59]. We can see that taking \(\lambda_{\pm}^2=0\) cannot be done for both signs since it will make both \(u\) and \(F\) trivial. Thus, we can only preserve half of the symmetry, hence the name, half-BPS states. In general, the Bogomolny equation from SUSY preservation \eqref{Equ} does not have to satisfy the on-shell equation. Substituting the on-shell equation \eqref{EqF} to the Bogomolnyi equations from trivial SUSY transformation gives
\begin{equation}\label{EqBPS}
    i\varepsilon_{jk}\partial_j u\partial_k \Bar{u}=\pm\frac{g}{2h},
\end{equation}
that is the on-shell BPS equation of the \(\mathcal{N}=2\) Baby Skyrme model. This equation is similar to the BPS equation of the non-supersymmetric Baby Skyrme model with the term on the right-hand side of \eqref{EqBPS}, which depends on the Kaehler metric, which resembles the potential term.

\subsection{The Novel Method by using SUSY extended BPS Lagrangian}
It is straightforward to deduce that from this standard method, we cannot produce a first-order equation of \(u\) for the non-BPS states. The non-BPS states can only be found from the second-order equations of \(u\), hence requiring an extra condition in order to solve it uniquely. Our newly proposed method can be used to alleviate such a problem in a more straightforward manner and the method can produce both BPS and non-BPS states of this example theory, as demonstrated in the two following sections. 

In this section, we are going to demonstrate how we can reproduce the BPS equations \eqref{EqBPS} mentioned above by using our new proposed method. 
If the BPS limit is achievable then we have \(\mathcal{L}^{\text{stat}}_{\text{T,bos}}=\mathcal{L}_{BPS}\). Let us use the following notation, \(\alpha=\partial_i u\partial^i \Bar{u}\) and \(\beta=i\varepsilon_{jk}\partial_j u\partial_k \Bar{u}\). For this case, the Lagrangian should be expressed as
\begin{eqnarray}
 \mathcal{L}^{\text{stat}}_{\text{T,bos}}&=&g\left[\alpha+F\Bar{F}\right]+h\left[(\alpha+F\Bar{F})^2-\beta^2\right]\
\end{eqnarray}
There are actually two ways to employ the BPS Lagrangian method in order to find the BPS equations from the Lagrangian above. The first one is done by simply completing the squares contained in the Lagrangian and the second one is done by finding the solutions of the BPS limit equation followed by calculating the constraint equations from the Euler-Lagrange equations of the BPS Lagrangian.

Let us consider a typical case for BPS Skyrmion. According to the BPS Lagrangian method [28], the minimal BPS Lagrangian consists of linear terms only, given by
\begin{equation}
    \mathcal{L}_{BPS}=-X_1\gamma-Y_1\beta,
\end{equation}
where we have defined \(\gamma=\alpha+F\Bar{F}\).
Again, by assuming that the BPS limit is satisfied, we have
\begin{equation}
    h\gamma^2+\left(X_1+g\right)\gamma-h\beta^2+Y_1\beta=0.
\end{equation}
If we solve for \(\gamma\) and \(\beta\) in order to find the corresponding Bogomolnyi equation, then we arrive at the following set of equations
\begin{equation}\label{BogHalf}
    \gamma=\mp\frac{Y_1}{2h},~~~\beta=\frac{Y_1}{2h}.
\end{equation}
Imposing the condition of single-solutions for both \(\gamma\) and \(\beta\) gives \(Y_1^2-(X_1+g)^2=0.\)
There are additional constraint equations which need to be considered from the Euler-Lagrange equations of the BPS lagrangian,
\begin{equation}
    \mathcal{L}_{BPS}=-(\pm Y_1-g)(\alpha+F\Bar{F})-Y_1 \beta.
\end{equation}
Variation of the above Lagrangian with respect to the auxiliary field gives \(\Bar{F}(\pm Y_1-g)=0\), implying that 
\begin{equation}
    Y_1=\pm g.
\end{equation}
By substituting this result back to the Bogomolnyi equations \eqref{BogHalf}, we managed to reproduce the on-shell BPS equations via the BPS Lagrangian method, namely
\begin{equation}
    i\varepsilon_{jk}\partial_j u\partial_k \Bar{u}=\pm\frac{g}{2h}.
\end{equation} 
Notice that the BPS equations obtained from the standard method coincide with the BPS equations obtained using this novel method based on the supersymmetric generalization of the BPS method. Furthermore, the on-shell equation for \(F\), \eqref{EqF}, is automatically incorporated as the Bogomolnyi equations \eqref{BogHalf}, hence the equations are guaranteed to be on-shell.

\section{Bogomolny Equations of the Non-BPS States  using the Novel Method}
In this section, we demonstrate how we can construct a set of first-order equations corresponding to the non-BPS states by using the BPS Lagrangian method. Consider the most general BPS Lagrangian for our model, namely
\begin{equation}
    \mathcal{L}_{BPS}=-W\beta\gamma-X_1\gamma-X_2\gamma^2-Y_1\beta-Y_2\beta^2-Z,
\end{equation}
where \(W,~ X_2,~ Y_2\) and \(Z\) are the additional auxiliary functions, the corresponding solutions should break more supersymmetries. 
The general Bogomolnyi equations from the quadratic completion are
\begin{equation}
    \gamma=-\frac{g+W\beta+X_1}{2(h+X_2)},~~~\beta=-\frac{W(X_1+g)-2(h+X_2)Y_1}{4h^2+W^2+4hX_2-4(h+X_2)Y_2}
\end{equation}
(note that we can also express \(\beta\) in terms of \(\gamma\) as an alternative equivalent way to express the Bogomolnyi equations) which results in the following equation for the remainder
\begin{eqnarray}
    Y_2 \left(g^2-4 h Z\right)+g^2 (-h)-X_1
   \left(2 g \left(h-Y_2\right)+W Y_1\right)-g
   W Y_1+4 h^2 Z&&\nonumber\\+X_2 \left(4 h Z-4 Y_2
   Z+Y_1^2\right)+X_1^2 \left(Y_2-h\right)+h
   Y_1^2+W^2 Z&=&0.
\end{eqnarray}
At this point, we need to impose some assumptions since we only have three possible constraints which came from the remainders and two equations from the Euler-Lagrange equations of \(\mathcal{L}_{BPS}\) with respect to \(F\) and \(u\). As such, we can only solve at most three unknowns from these constraints, which implies that the generalized \(\mathcal{L}_{BPS}\) can only have at most three terms. One of the simple cases that is straightforward to solve but still accommodates higher power terms is \(X_2=X_1=W=0\).
Such choice is inspired by the constraint equation from EL equations of \(\mathcal{L}_{BPS}\) with respect to \(F\), namely
\begin{equation}\label{consNonBPS}
    2X_2\gamma+W\beta+X_1=0.
\end{equation}
We can see that \(X_2=X_1=W=0\) trivially satisfies the constraint \eqref{consNonBPS}.
This implies that the remaining terms in the BPS Lagrangian are \(-Y_1\beta-Y_2\beta^2-Z.\)
Using this particular choice of \(X_2,~X_1,\) and \(W\), we can recast the BPS limit equation as
\begin{equation}\label{QuadNonBPS}
    h\gamma^2+g\gamma+\left(Y_2-h\right)\beta^2+Y_1\beta+Z=0.
\end{equation}
The Bogomolny equation for \(\gamma\) and  \(\beta\) are given by
\begin{eqnarray}\label{BogNonBPS1}
    \gamma=-\frac{g}{2h},&&\,\,\,\,\,\,\,\,\,\,\,\,
\label{BogNonBPS2}
\beta=\frac{Y_1}{2\left(h-Y_2\right)},
\end{eqnarray}
and from the remainder of \eqref{QuadNonBPS} we have \(
0=\left(Y_2-h\right)\left(4Zh-g^2\right)-hY_1^2.\)
This equation gives us the expression for \(Z\), namely
\begin{equation}\label{ZNonBPS}
    Z=\frac{1}{4h}\left[g^2+Y_1^2\left(\frac{h}{Y_2-h}\right)\right].
\end{equation}
This left us with two functions, \(Y_1\) and \(Y_2\), to be determined by the constraint equation from the variation of \(\mathcal{L}_{BPS}\) with respect to \(u\), hence gives us only one arbitrary functions that parameterize the family of solutions.

The BPS Lagrangian for this case is given by
\begin{equation}
    \mathcal{L}_{BPS}=-Y_2(\varepsilon_{jk}\partial_ju\partial_k\Bar{u})^2-iY_1\varepsilon_{jk}\partial_j u\partial_k \Bar{u}-Z,
\end{equation}
The variational principle with respect to \(\bar{u}\) leads to the constraint equation \(-iuZ'=Y_2\varepsilon_{jk}\partial_ju\partial_k\beta.\)
If we substitute the Bogomolnyi equations and the expression for \(Z\equiv Z(u\Bar{u})\) form equation \eqref{ZNonBPS}, then we have the differential equation for \(Y_1\) that depends on \(Y_2\), namely
\begin{eqnarray}
        0&=&\left(Y_2-h\right)^3\left(\frac{g^2}{h}\right)'+\left(Y_2'-h'\right)hY_1^2+\frac{(Y_2-h)(Y_2-2h)}{2}\left(Y_1^2\right)',
\end{eqnarray}
with \((.)'\) denotes the derivatives with respect to \(u\bar{u}\).
It is straightforward to demonstrate that half of the symmetry is preserved if and only if \(Y_2=0\) or \(Y_2=\frac{h}{g}\). The eigenvalues of supersymmetry transformation on the fermion fields in terms of functions \(g,~h,~Y_1\) and \(Y_2\) satisfies 
\begin{equation}
    \lambda_{\pm}^2=\frac{g}{2h}\mp\frac{Y_1}{2(Y_2-h)}.
\end{equation}
Suppose that half of the symmetry is preserved, then the only set of solutions of constraint equation \eqref{consNovBPS} is either \(Y_2=0\) or \(Y_2=\frac{h}{g}\). The opposite is obvious since if we choose \(Y_2=0\) then the set of equations for half-BPS states is recovered and if we choose \(Y_2=\frac{h}{g}\) then \(Y_1=\pm(1-g)\) is the solution of \eqref{consNovBPS} which leads to \(\lambda_{\pm}=0\). Since for general \(Y_2\) all of the supersymmetries are broken, then the solutions with \(Y_2\neq 0\) correspond to the non-BPS states.

%When \(Y_2\neq 0\) then we shall expect more broken symmetry. Let us consider the quarter(\(1/4\))-BPS states where three out of four fermion transformation is broken. \dots

Now, let us consider a special case where \(Y_2\) is linear to \(h\), namely \(Y_2=C_0+C_1 h\), and \(h\) satisfies the relation for standard Baby Skyrme model, \(h=g^2\), with the Kaehler metric is given by the standard baby Skyrme model, \(g=\frac{1}{(1+u\Bar{u})^2}\). Such states should, in general, break all supersymmetries. Thus, from equation \eqref{consNovBPS} we deduce that
\begin{equation}
    Y_1=\pm \sqrt{2}\frac{(1-C_1)h-C_0}{[(2-C_1)h-C_0]^{\frac{C_1-1}{C_1-2}}},
\end{equation}
and the constant \(C_0\) should be bounded from above by \((2-C_1)\min_{\{u,\bar{u}\}}h\) such that equation \eqref{ConstY2} is non-singular for all \(u,\Bar{u}\). Since the minimum of \(h\) is zero, then we can conclude that \(C_0\) is zero.
Thus, for this special case, the on-shell Bogomolnyi equation is given by
\begin{equation}\label{ConstY2}
    i\varepsilon_{jk}\partial_j u\partial_k \Bar{u}=\pm \frac{1}{\sqrt{2}}\left[\frac{(2-C_1)}{(1+u\bar{u})^4}\right]^{-\frac{C_1-1}{C_1-2}}.
\end{equation}
The Bogomolnyi equation shown above is connected to the BPS equation by taking the limit \(C_1=0\).

In general, solutions of \(u\) in closed form are not possible. In order to find the explicit solutions for \(u\), we need to assume the desired symmetries. There are a lot of possible symmetries for a static Skyrme model on \(\mathbb{R}^2\) but the most well-known are the spherically symmetric case. The spherically symmetric Skyrmions can be obtained if we assume that the solutions are in hedgehog form, namely \(u(r,\theta)=\xi(r)e^{if(\theta)}\) [30, 31, 34].
Substituting this form of solutions to \eqref{ConstY2} gives
\begin{equation}\label{effEq}
    \frac{2\xi}{r}\frac{d\xi}{dr} \frac{df}{d\theta}=\pm\frac{1}{\sqrt{2}}\left[\frac{(2-C_1)}{(1+\xi^2)^4}\right]^{-\frac{C_1-1}{C_1-2}},
\end{equation}
which implies that, through separation of variables, \(f=n\theta\) with \(n\in\mathbb{Z}\) such that \(u\) is periodic in \(\theta\rightarrow \theta+2\pi\). 

Let \(\mathcal{X}=\xi^2\) then the above equation can be recast into the differential equation for \(\mathcal{X}\).
It is straightforward to solve \eqref{effEq} and the solutions that satisfies \(u(0,\theta)=0\) is given by
\begin{eqnarray}
    \mathcal{X}&=&\left(1\pm(2-C_1)^{-2+\frac{1}{2-C_1}}(3C_1-2)\frac{r^2}{2\sqrt{2}n}\right)^{-\frac{C_1-2}{3C_1-2}}-1.\nonumber\\
    &&
\end{eqnarray}
We can see that the negative branch solutions are monotonically decreasing in \(r\) and are always negative. Since \(\mathcal{X}=u\Bar{u}\geq0\) then the negative branch should be neglected. Furthermore, there is a cut-off radius for the positive branch at 
\begin{equation}
    r_{\text{cut-off}}=\frac{2^{3/4} (2-C_1)^{\frac{1}{2}
   \left(2-\frac{1}{2-C_1}\right)}
   \sqrt{n}}{\sqrt{2-3 C_1}}
\end{equation} that are well defined for \(-2< C_1\leq0\), implying that the corresponding non-BPS solution is a compacton.

Because there are values of \(C_1\) that produce compact solutions, we shall consider whether they belong to compactons or not. In order to do so we need to study the distribution of the Skyrmion through its topological charge. The topological charge density is given by 
\begin{equation}
    q=\frac{n}{2\pi}\frac{\mathcal{X}'}{r(1+\mathcal{X})^2},
\end{equation}
with \(\mathcal{X}'=\frac{d\mathcal{X}}{dr}\) [29,34,59].
We found that for all physically allowed \( C_1\) the topological charge density, indeed, belongs to the compacton type because the topological charge goes to zero at the cut-off radius. 

It is important to note that the true field equations of the theory are the original equations from the Euler-Lagrange of \(\mathcal{L}^{\text{stat}}_{\text{T,bos}}\) and one should see the Euler-Lagrange equations of \(\mathcal{L}_{BPS}\) just as tools to add extra equations to constraint the solutions of the auxiliary functions from the zero discriminant condition mentioned previously. As such, the corresponding system of constraints and Bogomolnyi equations (first-order diff. eq.) should always satisfy the Euler-Lagrange equations of \(\mathcal{L}^{\text{stat}}_{\text{T,bos}}\) (second order diff. eq.) regardless of the choice of the \(\mathcal{L}_{BPS}\). This also means that the family of solutions to the Bogomolnyi equations are contained in the family of solutions to the original field equations of the theory.
In order to be sure that the BPS and non-BPS solutions found in our method do satisfy the original Euler-Lagrange equation, let us consider the on-shell field equation for \(u\) from the SUSY \(\mathcal{N}=2\) baby-Skyrme model, 
\begin{eqnarray}
    0&=&\left[\partial_i\left(\frac{\partial\mathcal{L}^{\text{stat}}_{\text{T,bos}}}{\partial~\partial_i \bar{u}}\right)-\frac{\partial\mathcal{L}^{\text{stat}}_{\text{T,bos}}}{\partial \bar{u}} \right]_{F\bar{F}=-\partial^iu\partial_i\bar{u}-g/2h}\nonumber\\
    \rightarrow0&=&uh'\left(i\varepsilon_{jk}\partial_j u\partial_k \bar{u}\right)^2+2hi\varepsilon_{jl}\partial_ju\partial_l\left(i\varepsilon_{mn}\partial_m u\partial_n \bar{u}\right)-u\left(\frac{g^2}{4h}\right)'.
\end{eqnarray}
For the half-BPS states, the Bogomolnyi equation is given by \(\beta=\pm g/2h\). Substituting this equation to the original field equation above gives us
\begin{eqnarray}
    0&=&uh'\left(\frac{g}{2h}\right)^2\pm 2hi\varepsilon_{jl}\partial_ju\partial_l\left(\frac{g}{2h}\right)-u\left(\frac{g^2}{4h}\right)'\nonumber\\
    &=&u\left[h'\left(\frac{g}{2h}\right)^2+g\left(\frac{g}{2h}\right)'-\left(\frac{g^2}{4h}\right)'\right]\nonumber\\
    &=&0,
\end{eqnarray}
which implies that the corresponding half-BPS solutions trivially satisfy the on-shell field equation of the SUSY \(\mathcal{N}=2\) baby-Skyrme model. We can also check the Bogomolnyi equation for the non-BPS state that is given by \(\beta=Y_1/2(h-Y_2)\) where \(Y_1\) and \(Y_2\) are related by equation (C7). Again, substituting the \(\beta\) from this non-BPS Bogomolnyi equation to the original field equation gives us
\begin{eqnarray}
    0&=&uh'\beta^2+ 2hi\varepsilon_{jl}\partial_ju\partial_l\beta-u\left(\frac{g^2}{4h}\right)'\nonumber\\
    &=&\frac{u}{4}\left[\frac{h'Y_1^2}{(h-Y_2)^2}+2\frac{hY_1}{(h-Y_2)}\left(\frac{Y_1}{(h-Y_2)}\right)'-\left(\frac{g^2}{h}\right)'\right].
\end{eqnarray}
We can rearrange the equation above into (C9) for non-trivial \(u\), implying that both are equivalent. As such, the original field equation is satisfied by \(\beta\), hence the solutions of the non-BPS Bogomolnyi equation do satisfy the original field equation which came from the Euler-Lagrange equation of the theory.

\section{Energy and Topologically Supported Stability of BPS and Non-BPS States}
The corresponding total energy for the BPS states corresponding to the BPS equation \eqref{EqBPS} can be calculated directly from the Hamiltonian density of static models, namely \(H=- \mathcal{L}^{\text{stat}}_{\text{T,bos}}\), provided the fermionic sector is zero due to trivial fermion fields. From the Hamiltonian density, we shall proceed by transforming \(u\) into the hedgehog ansatz by assuming that \(\mathcal{L}^{\text{stat}}_{\text{T,bos}}\) is spherically symmetric in \(\mathbb{R}^2\). This can be done by taking \(u=\xi(r)e^{i f(\theta)}\) and both \(g,h\) are functions of \(u\bar{u}\), such that the only possible solution for \(f(\theta)\) is \(f=n\theta\), where \(n\in\mathbb{Z}\), \(n\) is the topological charge of the Skyrmion [28, 39, 34]. Substituting the hedgehog ansatz and integrating \(H\) at the BPS limit on the whole space gives us
\begin{eqnarray}
    E&=&\int_0^\infty\int_0^{2\pi}\frac{\xi}{(1+\xi^2)^2}\frac{d\xi}{dr}\frac{df}{d\theta}~drd\theta\nonumber\\
    &=&n\pi.
\end{eqnarray}
It is implied that the energy of the BPS states is topologically protected, hence, such states are stable states.

The same argument of topologically protected stability can be applied to the non-BPS states as well. in order to do so, we need to check whether the total static energy is linear to the topological charge or not. To do this, we can employ the same approach by integrating the Hamiltonian density for the non-BPS solutions, namely
\begin{eqnarray}
    E&=&\int_0^\infty\int_0^{2\pi}\frac{\sqrt{2}(2-C_1)^{-\frac{C_1-1}{C_1-2}}\xi}{(1+\xi^2)^{\frac{4}{2-C_1}}}\frac{d\xi}{dr}\frac{df}{d\theta}~drd\nonumber\\
    &=&n\pi\sqrt{2}(2-C_1)^{-\frac{C_1-1}{C_1-2}}\frac{\Gamma\left(\frac{2+C_1}{2-C_1}\right)}{\Gamma\left(\frac{4}{2-C_1}\right)}.
\end{eqnarray}
We can see that the energy is linear in \(n\), implying that the BPS property \(E(n_1)+E(n_2)=E(n_1+n_2)\) is still preserved for these solutions although all the supersymmetries are broken. As such, these states are stable because they are topologically protected states unless \(n=0\). Furthermore, from the expression of total energy \eqref{energy} we know that \(C_1\in(-2,0]\) because, if \(C_1\in(-\infty,-2]\cup[2,\infty)\) the energy is either negative or undefined, and if \(C_1\in(0,2)\) the energy of the non-BPS solution is lower than the BPS bound.
%\section{Appendixes}

%To start the appendixes, use the \verb+\appendix+ command.

% The \nocite command causes all entries in a bibliography to be printed out
% whether or not they are actually referenced in the text. This is appropriate
% for the sample file to show the different styles of references, but authors
% most likely will not want to use it.
%\nocite{*}

\end{document}